\documentclass[runningheads]{llncs}
\usepackage[T1]{fontenc}
\usepackage{amsmath}
\usepackage{booktabs}  % For \toprule, \midrule, \bottomrule
\usepackage{array}
\usepackage{hyperref}
\usepackage{listings}
\usepackage{graphicx}
\usepackage{todonotes}

\begin{document}
\title{SIDEKICK: A Semantically Integrated Resource for Drug Effects, Indications, and Contraindications}
\titlerunning{SIDEKICK}
\author{Mohammad Ashhad\inst{1} \and
Olga Mashkova\inst{2} \and
Ricardo Henao\inst{3} \and
Robert Hoehndorf \inst{2}
}
\authorrunning{M. Ashhad et al.}
\institute{BESE, King Abdullah University of Science and Technology, Thuwal, KSA \and
CEMSE, King Abdullah University of Science and Technology, Thuwal, KSA \and
Dept. of Bioinformatics and Biostatistics, Duke University, Durham, USA
\email{firstname.lastname@kaust.edu.sa/duke.edu}\
% \url{http://www.springer.com/gp/computer-science/lncs} 
}
\maketitle              % typeset the header of the contribution
\begin{abstract}
  Pharmacovigilance and clinical decision support systems utilize
  structured drug safety data to guide medical practice. However,
  existing datasets frequently depend on terminologies such as MedDRA,
  which limits their semantic reasoning capabilities and their
  interoperability with Semantic Web ontologies and knowledge
  graphs. To address this gap, we developed SIDEKICK, a knowledge
  graph that standardizes drug indications, contraindications, and
  adverse reactions from FDA Structured Product Labels. We developed
  and used a workflow based on Large Language Model (LLM) extraction
  and Graph-Retrieval Augmented Generation (Graph RAG) for ontology
  mapping. We processed over 50,000 drug labels and mapped terms to
  the Human Phenotype Ontology (HPO), the MONDO Disease Ontology, and
  RxNorm. Our semantically integrated resource outperforms the SIDER
  and ONSIDES databases when applied to the task of drug repurposing
  by side effect similarity. We serialized the dataset as a Resource
  Description Framework (RDF) graph and employed the Semanticscience
  Integrated Ontology (SIO) as upper level ontology to further improve
  interoperability. Consequently, SIDEKICK enables automated safety
  surveillance and phenotype-based similarity analysis for drug
  repurposing.

\end{abstract}
\textbf{Resource type}: Knowledge Graph \\
\textbf{License}: CC BY 4.0 International \\
\textbf{DOI}: https://doi.org/10.5281/zenodo.17779317 \\
\textbf{URL}: http://sidekick.bio2vec.net 

\section{Introduction}

Drug safety information is fundamental to clinical decision-making,
yet accessing this knowledge and integrating it with other clinical
information remains a challenge. Adverse drug reactions (ADRs)
represent a leading cause of morbidity and mortality, with medication
errors estimated to cost \$42 billion annually in the United
States~\cite{classen1997adverse}. Despite their clinical importance,
the systematic organization and computational representation of drug
indications, contraindications, and side effects has lagged behind
other areas of biomedical informatics.

Several existing drug safety datasets have contributed to
pharmacovigilance research. SIDER~\cite{kuhn2016sider} was the first
open database to systematically extract adverse drug reactions from
drug labels, mapping both side effects and indications to the Medical
Dictionary for Regulatory Activities (MedDRA)~\cite{brown2004meddra}.
However, SIDER's indication data were extracted primarily as a
byproduct to filter false positives during adverse event
extraction. OnSIDES~\cite{tanaka2025onsides} employed fine-tuned
PubMedBERT models to extract over 3.6 million drug-adverse event pairs
from FDA labels but focused exclusively on adverse events without
extracting therapeutic indications or contraindications; OnSIDES also
uses MedDRA for standardization. OFFSIDES and
TWOSIDES~\cite{tatonetti2012offsides}, while important for
pharmacovigilance, employ a different method by mining 
post-marketing surveillance data from the FDA adverse event reporting system (FAERS) to identify off-label effects and drug--drug interactions using statistical techniques. Notably, no existing resource provides
comprehensive coverage of contraindications that are critical safety
information that indicates when drugs should not be used.

Although these resources are widely used in biomedical informatics
research, they share a common limitation: reliance on MedDRA for the
standardization of drug effects and indications. Although MedDRA is
widely adopted in a regulatory context, MedDRA imposes constraints on
computational interoperability due to its architectural design.
Fundamentally, MedDRA functions as a terminology rather than an
ontology; terminologies primarily facilitate data access and indexing,
while ontologies exist to structure the underlying domain through
explicit semantic relationships and axioms~\cite{guarino1998formal}.

Drug repurposing, which seeks to identify new therapeutic uses for
existing medications, increasingly relies on automated reasoning about
drug--disease relations~\cite{pushpakom2019repurposing}, and
phenotype-based approaches to drug repurposing use the observation
that drugs with similar side effects likely share mechanisms of
action~\cite{campillos2008drug}. Such analyzes benefit from detailed
phenotypic descriptors and the ability to compute semantic similarity
between side effects, and semantic similarity measures improve with
the ability to deductively infer relations between
terms\cite{kulmanov2021semantic,pesquita2009semantic}.

We developed SIDEKICK to address the limitations of existing resources
focused on drug safety. Our contributions are: (1) We extended the
clinical scope beyond side effects and integrated indications,
contraindications, and adverse reactions from over 50,000 FDA
Structured Product Labels; (2) We used an extraction method based on
Large Language Models combined with Graph-RAG to improve both
granularity and coverage of extracted information; (3) We normalized
clinical entities to standardized ontologies, mapping side effects and
phenotypic contraindications to the Human Phenotype Ontology (HPO)
\cite{kohler2021human}, 
% \rh{[missing ref]}, 
disease terms to the MONDO Disease Ontology \cite{vasilevsky2022mondo}, 
% \rh{[missing ref]}, 
and drug entities to RxNorm; (4) We serialized SIDEKICK as a Resource
Description Framework (RDF) \cite{pan2009resource} graph using the Semanticscience
Integrated Ontology (SIO) \cite{dumontier2014sio} as an upper-level framework, further
improving interoperability and adherence to FAIR data principles
\cite{wilkinson2016fair};
% \rh{[missing ref], also the sentence is way to long and difficult to follow}; 
and (5) We evaluated SIDEKICK against the OnSIDES \cite{tanaka2025onsides} 
% \rh{[missing ref]}
baseline to demonstrate that SIDEKICK improves the prediction of shared
drug targets by side effect similarity.

\section{Related work}

\subsection{Drug safety databases and their applications}
SIDER~\cite{kuhn2016sider} established the precedent for extracting
adverse drug reactions (ADRs) from package inserts, providing
frequency data and mappings to the Medical Dictionary for Regulatory
Activities (MedDRA) \cite{brown1999medical}.
% \rh{[missing ref]}. 
% The resource also includes a dataset of drug indications; however, these were extracted primarily to reduce
% false positives during ADR extraction by identifying medical terms
% that do not correspond to adverse events, rather than serving as
% authoritative therapeutic use data 
The resource also includes a dataset of drug indications, though these should not be considered authoritative therapeutic use data. They were extracted primarily to reduce false positives during ADR extraction by helping to identify medical terms not corresponding to adverse events.
% \rh{weird sentence}. 
Most recently, OnSIDES~\cite{tanaka2025onsides} employed fine-tuned PubMedBERT models
to extract over 3.6 million drug-adverse event pairs from FDA labels,
achieving high extraction performance. OnSIDES
uses MedDRA for standardization and focuses exclusively on adverse
drug events; it does not extract therapeutic indications or
contraindications.

Despite these advancements, these resources universally employ MedDRA
for standardization. MedDRA functions as a terminology designed to
facilitate regulatory data access rather than a formal ontology
designed to structure the domain. Consequently, MedDRA aggregates
terms based on retrieval expectations rather than shared biological
properties or logical definitions. Consequently, MedDRA's
hierarchical structure may aggregate ontologically distinct
terms. Furthermore, this terminology lacks explicit logical linkage
to other types of biological entities, such as anatomical structures
defined in Uberon~\cite{mungall2012uberon} or physiological processes
defined in the Gene Ontology (GO)~\cite{gene2019gene}. In contrast,
phenotype ontologies such as the Human Phenotype Ontology (HPO) utilize
logical definitions based on the Entity-Quality (EQ) design
pattern~\cite{mungall2010integrating,gkoutos2018anatomy}.
%\rh{[missing ref]}. 
This axiomatic structure of phenotype ontologies therefore enables the
semantic interoperability required to link clinical phenotypes with
their underlying biological mechanisms, a capability that the
access-oriented architecture of MedDRA does not support.

\subsection{Phenotype and disease ontologies}
To overcome the limitations of using terminologies, modern biomedical
informatics employs formal ontologies that support deductive reasoning
\cite{hoehndorf2015role}.
%\todo{The role of ontologies in biological and biomedical research. Briefings in Bioinformatics, 2017(?)}. 
The Human Phenotype Ontology (HPO)~\cite{kohler2021human} provides a standardized vocabulary of phenotypic abnormalities organized as an axiomatic
theory (a formal ontology) using the Web Ontology Language (OWL)
\cite{antoniou2009web}.  HPO enables the calculation of semantic
similarity between terms by exploiting not only a graph structure but
also its axioms \cite{smaili2020formal}.
% \rh{[missing ref]}. 
The axioms are important for identifying
drugs with similar side effect profiles, where similarity may be based
on shared mechanisms of action or shared anatomical structures.
Similarly, the MONDO Disease Ontology~\cite{vasilevsky2022mondo}
harmonizes disease definitions across different databases % (OMIM,
% Orphanet, NCIT) 
using equivalence axioms. By mapping clinical entities to HPO and
MONDO, we enable the application of algorithms that require deductive
reasoning and axiomatic theories, a capability absent in current
MedDRA-based resources.

\subsection{Semantic interoperability and upper-level ontology}
Semantic interoperability across heterogeneous biomedical datasets
requires more than standardized vocabularies; it also requires a
structural framework to define relationships between entities.
Upper-level ontologies provide such a framework by defining general
classes ({\em e.g.}, processes, objects, attributes) that remain consistent
across domains and providing relationships that can be used across
resources. The Semanticscience Integrated Ontology
(SIO)~\cite{dumontier2014sio} serves as a lightweight upper-level
ontology widely adopted in scientific applications of the Semantic
Web.

Upper-level ontologies also provide Ontology Design Patterns (ODPs)
\cite{gangemi2009ontology} --- reusable modeling solutions for
recurrent ontology design problems. For example, SIO employs ODPs to
model associations as reified entities ({\em i.e.}, classes). This practice
allows for the attachment of metadata (such as provenance or evidence
strength) directly to the association. Consequently, by utilizing
classes and relations from SIO, we improve interoperability because
the use of an upper level ontology allows users and agents to query the
knowledge graph using standard SIO-based query patterns, and without
requiring explicit knowledge of how each interaction is specifically
represented. This approach contrasts with ad-hoc schema designs that
may be found in standalone databases, which limit data federation and
integration.

\section{Methodology}
\subsection{Overview}
\begin{figure*}[t]
    \includegraphics[width=\textwidth]{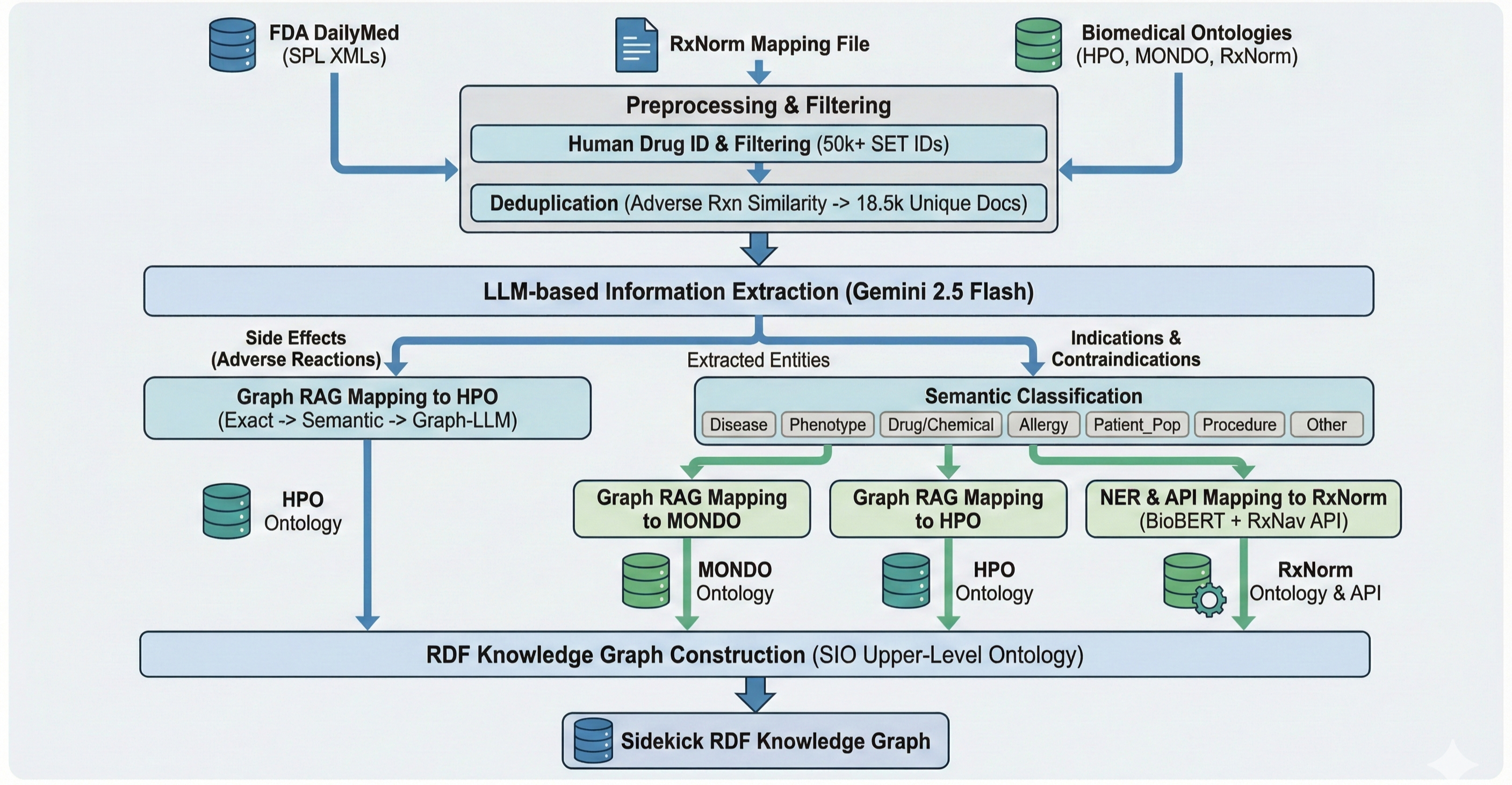}
    \caption{Workflow for creating the SIDEKICK knowledge graph.}
    \label{fig:methodology}
\end{figure*}

We developed a workflow to extract, structure, and map drug-related
clinical information from FDA Structured Product Labels (SPLs) to
standardized biomedical ontologies. The workflow consists of seven
main stages: (1) identification and filtering of human prescription
drugs, (2) deduplication of redundant SPL documents, (3) automated
extraction of clinical entities using large language models (LLMs),
(4) mapping of side effects to the Human Phenotype Ontology (HPO), (5)
classification of contraindications and indications, (6) mapping of
disease and phenotype terms to MONDO and HPO ontologies, respectively,
and (7) mapping of drug--drug interaction terms to RxNorm. Finally,
all extracted and mapped information is integrated into an RDF
knowledge graph. Figure~\ref{fig:methodology} illustrates the full
workflow.

\subsection{Data collection and preprocessing}
We acquired drug labeling data from the FDA DailyMed
database\footnote{\url{https://dailymed.nlm.nih.gov/}}, which serves
as the official repository for FDA-approved package inserts. These
documents adhere to the Health Level Seven (HL7) \cite{quinn1999hl7}
Structured Product Labeling (SPL) standard
\footnote{\url{https://www.fda.gov/industry/fda-data-standards-advisory-board/structured-product-labeling-resources}},
an XML-based format used for the exchange of product information.  To
isolate relevant data, we processed the DailyMed human prescription
label archives to extract the SPL Set Identifier (SET ID), a globally
unique identifier that persistently represents a specific labeling
document version. This process yielded 50,559 unique SET IDs
corresponding to human prescription drugs. Subsequently, we
standardized these entries by linking the SPL documents to RxNorm, the
normalized naming system for clinical drugs produced by the National
Library of Medicine (NLM). Using the NLM mapping files, we associated
each document with its specific RxNorm Unique Identifier
(RxCUI). Finally, to map branded products back to their fundamental
bioactive components, we queried the RxNav REST API; for each product
RxCUI, we retrieved the related terms and filtered strictly for active
ingredients (term type `IN').

For standardized term mapping, we utilized three major biomedical
ontologies: the Human Phenotype Ontology (HPO) \cite{kohler2021human},
which provides standardized vocabulary of phenotypic abnormalities with
over 16,000 terms; the MONDO Disease Ontology
\cite{vasilevsky2022mondo}, which integrates disease terminologies
from multiple sources; and RxNorm, accessed via the RxNav REST
API\footnote{\url{https://rxnav.nlm.nih.gov/}} for normalized clinical
drug names.

To mitigate computational overhead, we implemented a deduplication
strategy centered on the textual similarity of adverse reaction
profiles. First, we extracted the ``Adverse Reactions'' sections from
the SPL documents, identifying them via their specific Logical
Observation Identifiers Names and Codes (LOINC) identifiers.  Within
each grouping defined by a Product RxCUI, we performed pairwise
comparisons of these sections.  We used a two-tiered matching
approach: (1) exact content matching utilizing MD5 cryptographic
hashes to identify identical text; and (2) fuzzy matching utilizing
the SequenceMatcher algorithm \cite{ratcliff1988pattern} to identify
near-duplicates with a similarity threshold of 95\%. For every
group of similar documents, we retained the first encountered SET ID
as the representative entry. This filtering process reduced the
corpus from 50,559 to 18,500 unique documents ({\em i.e.}, 63.4\% reduction).

\subsection{Information extraction from SPL documents}
\subsubsection{Text preprocessing}
Prior to LLM-based extraction, we preprocessed SPL XML documents to
create structured text suitable for language model processing. We
excluded sections not relevant to clinical decision-making using a
blacklist of LOINC section codes, 
% \todo{add list to appendix}, 
including package labeling (\texttt{51945-4}), product data elements (\texttt{48780-1}), {\em etc}. For
retained sections, we preserved the hierarchical structure with
section titles, body text, and tables. We converted tables to a text
representation with rows separated by newlines and cells separated by
pipe characters.

\subsubsection{Large Language Model extraction}
We used the Google Gemini 2.5 Flash model via the OpenRouter API for
automated extraction of clinical entities from processed SPL text. The
model was prompted to extract three categories of information: (1)
Indications: medical conditions, diseases, or symptoms for which the
drug is indicated; (2) Contraindications: conditions, situations, or
patient populations in which the drug should not be used; (3) Adverse
reactions (side effects): undesirable effects associated with drug
use. The extraction prompt
% \todo{if possible, add the prompt as appendix}
instructed the model to search over all sections (not just those
with dedicated section headers), extract items in concise form while
maintaining accuracy. The model returned structured XML output with
nested tags for each entity category. We processed the deduplicated
SPL documents in batches of 500 with 10-second sleep intervals to
adhere to the API rate limits.

\subsection{Ontology mapping}
After extracting clinical entities from SPL documents, we mapped the terms
to standardized biomedical ontologies to enable interoperability and
integration with existing knowledge bases.
%\subsubsection{Side Effects Mapping to HPO} 
We then mapped the extracted adverse reactions to the Human Phenotype Ontology
(HPO) \cite{kohler2021human} using a Graph-Retrieval Augmented
Generation (Graph RAG) approach. We loaded the HPO ontology into the OBO
format
\footnote{\url{https://owlcollab.github.io/oboformat/doc/obo-syntax.html}}
and constructed a NetworkX \cite{hagberg2008exploring} directed graph
with a bidirectional dictionary that maps the term names and synonyms to
HPO IDs. We computed dense vector embeddings for all HPO terms and
synonyms using the \texttt{all-MiniLM-L6-v2} sentence transformer
model, processing in batches of 1,000 terms and caching to disk. The
embedding matrix contained representations for over 45,000 terms and
synonyms.

For each side effect term, we used a three-stage matching
strategy. First, we used exact string matching against primary term
names and synonyms (case-insensitive). Second, we used a semantic
similarity search based on the cosine similarity against all
embeddings, retrieving the top-10 matches and deduplicated by HPO
ID. In the third step, we used graph-enhanced LLM mapping where the
top semantic matches served as seed nodes for graph traversal to
identify related classes (parents, children, siblings), limited to 15
nodes. We then constructed new prompts for Gemini 2.5 Flash that contained
the top 5 semantic matches with similarity scores, definitions,
related terms with relationship annotations, and instructions to
return HPO ID and term name in JSON format. The responses were validated
to ensure that the returned IDs exist and that the term names match canonical names,
with automatic retries (up to 3 attempts). Unmappable terms were
assigned to \texttt{HP:0000001} (``All'', the top-level class of the
HPO).

%\subsubsection{Mapping indications and contraindications} 
We implemented an ontology-based mapping workflow to handle the
heterogeneous nature of contraindication and indication terms.
%\paragraph{Entity Classification }
First, we classified terms that occur in the indication and
contraindication sections of the SPL into seven categories: {\em
  Disease} (medical conditions, disorders, syndromes), {\em Phenotype}
(observable clinical signs, symptoms), {\em Drug or Chemical} (drug
interactions, concomitant medications), {\em Allergy or
  Hypersensitivity} (allergic reactions), {\em Patient Population}
(demographics, life stages), {\em Procedure} (medical/surgical
procedures), and {\em Other} (which corresponds to any other type for
which we could not identify a specific category). We implemented a
two-stage classification: (1) keyword-based auto-classification for
allergy-related terms containing the strings ``hypersensitivity'',
``allergic'', or ``anaphylaxis''; (2) LLM-based classification using
Gemini 2.5 Flash Lite with terms batched in groups of 15, receiving
detailed category definitions and returning JSON-formatted
classifications.

%\paragraph{Category-specific mapping}

% \textbf{Disease terms to MONDO}: 
We mapped the terms classified as {\em Disease} to the MONDO Disease
Ontology using the same Graph RAG approach as we used for HPO
mapping. We constructed a NetworkX graph from the MONDO OBO file and
generated embeddings using \texttt{all-MiniLM-L6-v2}. The MONDO
ontology contains over 25,000 disease classes. We applied the
three-stage matching strategy (exact match, semantic similarity with
top-10 retrieval, graph-enhanced LLM mapping with Gemini 2.5 Flash),
with validation and retries. Unmappable terms defaulted to
\texttt{MONDO:0000001} ({\em Disease or disorder}).

% \textbf{Phenotype terms to HPO}: 
Terms classified as {\em Phenotype} were mapped to HPO using the
identical Graph RAG pipeline described above for the mapping of side effects.
We employed BioBERT-based chemical NER~\footnote{\url{https://github.com/librairy/bio-ner}}
(\texttt{alvaroalon2/biobert\_chemical\_ner})~\cite{alonso2021named}
% \todo{cite
%   https://oa.upm.es/67933/ and link to the Bio-NLP repo at
%   https://github.com/librairy/bio-ner} 
  to identify drug entities
within text, with fallback heuristic parsing to remove phrases like
``coadministration with'' and splitting on delimiters. For each
entity, we queried the RxNav REST API using: (1) exact match via
\texttt{/rxcui.json}, and (2) approximate match via
\texttt{/approximateTerm.json} (top 10 candidates) if exact match
failed. Terms without recognized entities or ingredient-level RxCUIs
were assigned a group labeled ``other''.

\section{Knowledge graph construction and validation}
\subsection{Data model and schema}
\begin{figure*}[t]
    \includegraphics[width=\textwidth]{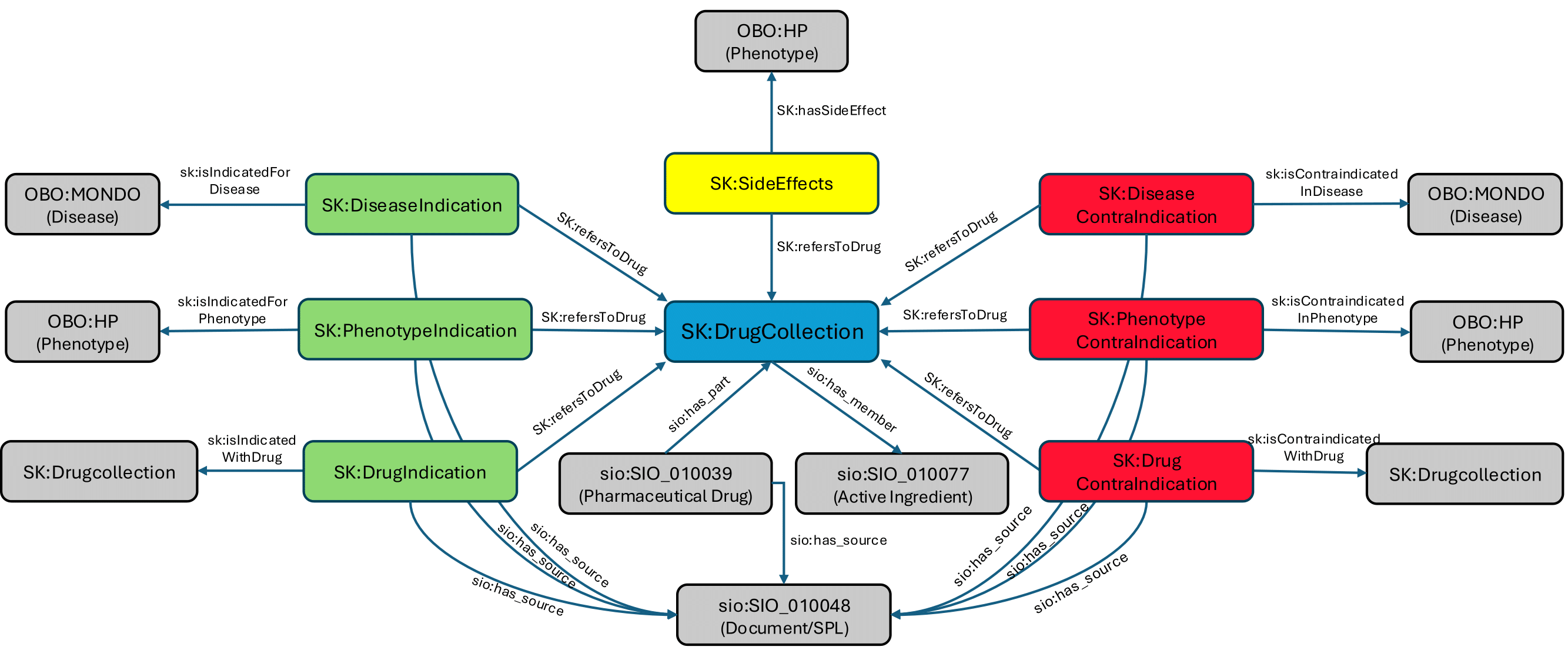}
    \caption{RDF schema for the SIDEKICK knowledge graph.}
    \label{fig:schema}
\end{figure*}

We constructed the SIDEKICK knowledge graph (KG) using the Resource
Description Framework (RDF) \cite{pan2009resource}. We selected the Semanticscience
Integrated Ontology (SIO) \cite{dumontier2014sio} as the upper-level
ontology due to its wide use in similar resources in the biomedical
domain. Figure~\ref{fig:schema} illustrates the schema that we developed for
the SIDEKICK knowledge graph.

%\subsubsection{Drug Representation}
We distinguish between the pharmaceutical product and the collection
of active ingredients it contains.
% \begin{itemize}
% \item 
A \textit{Pharmaceutical Drug} is a specific branded or generic
product (e.g., ``Naproxen 500 MG Oral Tablet''), which are typed as
\textit{PharmaceuticalDrug} (\verb_SIO:010039_) and identified by their Product RxCUI
using the Internationalized Resource Identifiers (IRIs) for the RxNorm
version provided by the NCBO BioPortal \cite{noy2009bioportal}.

% \item \textbf{Drug collection:} 
We introduced a new class, \textit{DrugCollection} (\verb_sk:DrugCollection_), which
is a subclass of \textit{Collection} (\verb_SIO:000616_), to represent the set of
active ingredients that are independent of the manufacturer or specific
formulation. For instance, ``Amoxicillin/Clavulanate'' as a drug
collection (\verb|sk:ingredient_set_723_48203|) represents the
combination of amoxicillin and clavulanate as active ingredients,
regardless of the brand name. This abstraction is necessary for three
reasons: (1) pharmaceutical products may contain multiple active
ingredients, (2) clinical relationships such as side effects and
therapeutic indications apply to the ingredient combination rather
than specific branded products, and (3) it prevents redundant
assertions across multiple branded or generic versions of the same
drug formulation. In practice, most drug collections are singleton
sets containing a single active ingredient, but the uniform
representation allows consistent handling of both single-ingredient
and combination products. Finally, individual active ingredients
are instances of {\em Active ingredient} (\verb_SIO:010077_) and use the
RxNorm IRIs from BioPortal.

The relationships are formalized such that a product \textit{has part} (\verb_SIO:000028_)
% (sio:has\_part\todo{is this really the name of the relation? It should
%   be the ID, not the name}, SIO:000028) 
a \textit{Drug Collection}, which in turn \textit{has member} (\verb_SIO:000059_) % (sio:has\_member\todo{same}, SIO:000059)
specific ingredients. We also assert a property chain, {\em 'has part'
  $\circ$ 'has member' $\subseteq$ 'has part'}. 

%\subsubsection{Association Reification}
We modeled clinical relationships (indications, contraindications, and
side effects) as reified 
% \rh{what is this word? --- ``reification'' is
%   the process of making a relation a class; ``reified'' is the
%   corresponding very.} 
associations (where relationships are represented as explicit entities) rather than simple edges. This
approach, consistent with a corresponding SIO design pattern, treats
the relationship itself as a node (an instance of \textit{Association},
\verb_SIO:000897_). We use this reified class for adding provenance, 
% \rh{how
%   so? --- using a 'has source' relation, which is to the association}
specifically the source SPL document, directly to the assertion via \textit{has source} relation (\verb_SIO:000253_).

To distinguish the two main participants in each clinical
relationship, we use two distinct properties: \textit{refers to drug}
(\verb_sk:refersToDrug_), a subproperty of \textit{refers to} (\verb_SIO:000628_),
% (sio:refers\_to, SIO:000628), 
always points to the drug collection being described (the
subject of the clinical finding), while specialized properties such as
\verb_sk:hasSideEffect_, \verb_sk:isIndicatedForDisease_,
% \todo{formatting; find a
%   consistent, non-obtrusive (not bold) way to format all relations;
%   all identifiers; all class.}, 
etc., point to the clinical entity
(phenotype, disease or another drug collection) that is the target of
the relationship. % All specialized properties are also subproperties of
% sio:refers\_to.

We defined seven specific subclasses of \textit{Association} (\verb_SIO:000897_) to
capture drug associations with their side effects, their indications
and contraindications (either diseases or phenotypes), other drugs to
be used in combination or to be avoided (a drug contraindication).
Provenance is captured using the \textit{has source} property (\verb_SIO:000253_), 
% (sio:has\_source, SIO:000253), 
which links each association node directly to the
specific SPL document node from which the information was extracted.
% This approach allows tracking of assertion 
% origins and enables handling of conflicting information across different SPL 
% versions or manufacturers. Every association node in SIDEKICK, regardless of type, 
% includes this provenance link, ensuring that users can trace any clinical claim 
% back to its authoritative FDA source.

\subsection{Triple generation workflow}
The graph generation workflow was implemented in Python using rdflib
\cite{krech2002rdflib}. We established Internationalized Resource
Identifiers (IRIs) in the SIDEKICK namespace for drug collections,
associations, and documents. For phenotypes and diseases, we use
identifiers in the OBO namespace, for drugs we use the 
namespace used in the BioPortal version of RxNorm, and for upper-level
classes and relations we use the SIO IRIs.

We use the Dublin Core \cite{baker2005maintaining} vocabulary for additional
metadata, in particular \verb_dcterms:description_ for textual descriptions
of classes and properties, \verb_dcterms:created_ for a creation timestamp,
\verb_dcterms:creator_ to specify author information, and \verb_dcterms:license_ to
provide licensing information. Additionally, we created a VOID file that provides basic metadata of the entire dataset.

The final SIDEKICK knowledge graph is serialized in Turtle format and
can be converted to any RDF
syntax. Table~\ref{tab:sidekick_statistics} summarizes basic
statistics of the SIDEKICK knowledge graph. The resulting graph
structure enables SPARQL queries across SIDEKICK and the referenced
ontologies. Beyond retrieval of directly asserted relationships (such
as identifying all drugs contraindicated for a specific phenotype),
the RDF structure enables reasoning over the HPO and MONDO ontology
hierarchies. For example, a SPARQL query can retrieve all drugs with
cardiac-related side effects by querying for phenotypes that are
subclasses of abnormality of the cardiovascular system (\verb_HP:0001626_) or
anatomically associated with heart structures via HPO's axioms that
use the UBERON anatomy ontology \cite{mungall2012uberon}.

\begin{table}[t]
\centering
\small
\caption{SIDEKICK knowledge graph statistics.}
\label{tab:sidekick_statistics}
\begin{tabular}{lr}
\toprule
\textbf{Component} & \textbf{Count} \\
\midrule
\multicolumn{2}{l}{\textit{Graph Structure}} \\
\quad Total RDF Triples & 3,155,094 \\
\quad Drug Collections (Hubs) & 3,036 \\
\quad Active Ingredients & 2,370 \\
\quad Pharmaceutical Products & 14,689 \\
\quad SPL Source Documents & 48,458 \\
\midrule
\multicolumn{2}{l}{\textit{Clinical Associations}} \\
\quad Total Associations & 205,683 \\
\quad Side Effects (Drug$\rightarrow$Phenotype) & 184,512 \\
\quad Disease Indications (Drug$\rightarrow$Disease) & 6,478 \\
\quad Phenotype Indications (Drug$\rightarrow$Phenotype) & 1,748 \\
\quad Drug Indications (Drug$\rightarrow$Drug) & 71 \\
\quad Disease Contraindications & 5,891 \\
\quad Phenotype Contraindications & 4,047 \\
\quad Drug Contraindications & 2,936 \\
\midrule
\multicolumn{2}{l}{\textit{Ontology Coverage}} \\
\quad Unique HPO Terms & 6,243 \\
\quad Unique MONDO Terms & 2,168 \\
\quad Unique RxNorm Ingredients & 2,370 \\
\bottomrule
\end{tabular}
\end{table}

\subsection{Schema and semantic validation}

To validate whether the SIDEKICK knowledge graph complies with the
schema in Figure~\ref{fig:schema}, we formalized the schema, and
additional constraints, using Shape Expressions (ShEx)~\cite{prud2014shape}.
% , a W3C standard
% for RDF schema validation. 
We validated all 205,683 clinical associations with respect to rules
that ensure that SIDEKICK (1) maintains structural integrity with
proper entity relationships, (2) include complete provenance linking
to FDA SPL source documents, (3) enforce ontology type safety with
pattern-matched IRIs ensuring correct use of MONDO and HPO
terminologies, and (4) satisfy all cardinality constraints. % The ShEx
% schema and validation code are publicly available at \url{}.
We further validated that our data is semantically consistent using
the ELK reasoner~\cite{kazakov2014incredible} (v0.5.0). Although SIO
contains axioms that exceed the OWL 2 EL profile \cite{krotzsch2012owl}, we were only
able to validate the EL subset due to the computational complexity of
OWL 2 DL reasoning \cite{krotzsch2012owl}. At release time, SIDEKICK is both consistent with respect to the
schema, and semantically consistent based on ELK.

% The validation process achieved 100\% compliance across all validation checks with zero structural or semantic errors detected.
% We implemented structural
% validation\todo{THIS should be done by ShEX.} during
% the generation process to ensure referential integrity (all subject
% and object IRIs are validated to ensure that referenced entities exist
% in the external vocabularies) and have types consistent with the
% identified association categories ({\em e.g.}, $DiseaseIndication$). We also
% implemented a filtering of assertions to top-level classes in
% ontologies to avoid uninformative assertions, and implemented a step
% to validate this \rh{unclear sentence}.

% The automated validation
% suite
% \todo{Rewrite this please; using Protege is not ``automated'';
%   and it is best to explain the rules and just validate with ShEX.}
% processed all 3,155,094 RDF triples, achieving 100\% compliance across all validation checks with zero structural or semantic errors detected.

% \textbf{Formal Schema Specification.}  We primarily formalized these constraints as a Shape Expressions
% (ShEx) schema~\cite{staworko2015complexity}, providing a
% machine-readable specification for automated validation of SIDEKICK
% and future updates.
% \todo{Provide validation results here; use pyshex
%   or shex-java, as you prefer}

\section{Experiments}
We conducted three evaluations to validate the quality and utility of
SIDEKICK: (1) human expert evaluation of mapping accuracy, (2)
prediction of shared drug targets using side effect similarity, and
(3) a competency question-based evaluation to demonstrate the
reasoning capabilities of the ontology-driven schema.

\subsection{Mapping quality evaluation}
%
%\subsubsection{Experimental setup}
%
To assess the accuracy of our Graph RAG mapping methodology, we
performed a human expert evaluation study. We randomly selected 50
unique SPL SET IDs from our dataset and extracted all unique side
effects (identified by HPO ID) from these documents, producing 980
distinct side effect mappings. Two domain experts with backgrounds in
pharmacology and biomedical informatics independently reviewed each
mapping. For each side effect, experts evaluated whether the assigned
HPO term accurately represented the original text from the drug
label. Mappings where the side effect name and HPO term matched
exactly were automatically validated without manual review. In cases
of disagreement between experts, they collaboratively discussed the
mapping until reaching consensus. Experts marked mappings as correct
if the HPO term was either an exact match or, when the original term
did not exist in HPO, represented the closest semantically appropriate
class available in the ontology.

%\subsubsection{Results}
Of the 980 mappings being evaluated, 948 were judged correct (overall
accuracy of \textbf{96.7\%}, 32 incorrect mappings). This accuracy
demonstrates the effectiveness of our Graph RAG methodology.  13 of
the 32 incorrect mappings involved terms with multiple possible
interpretations, requiring disambiguation. 10 errors occurred with
newly coined terms or drug-specific terminology not represented in
HPO.

\subsection{Drug target prediction via side effect similarity}
%
%\subsubsection{Motivation}
%
Drugs with similar side effect profiles often share molecular
targets~\cite{campillos2008drug,nugent2016computational}. We
hypothesized that SIDEKICK's ontology-based representation would
better capture mechanistic similarity compared to existing
vocabularies, enabling more accurate prediction of shared drug
targets.

%\subsubsection{Experimental Setup}
%
We compared SIDEKICK against OnSIDES~\cite{tanaka2025onsides}, the
current state-of-the-art drug side effect database. OnSIDES has been
previously shown to outperform SIDER~\cite{kuhn2016sider} on the same
task; therefore, comparison to OnSIDES represents comparison against
the best available resource.

% \paragraph{Ground Truth} 
We used DrugBank 5.1.12~\cite{wishart2018drugbank} as the ground truth
for drug--target associations. For each drug, we extracted all protein
targets and created drug pairs. Positive pairs were defined as drugs
that share at least one protein target; negative pairs were drugs with
no shared targets.

% \paragraph{Dataset Matching} 
To ensure a fair comparison, we identified drugs present in all three
datasets: SIDEKICK, OnSIDES (US labels only), and DrugBank. This
yielded 1,105 matched drugs, resulting in 13,983 positive pairs
(shared targets) and 595,977 negative pairs (no shared targets).
% \paragraph{Ontology-Based Similarity} 
We applied identical semantic similarity measures to both
datasets. Both SIDEKICK (HPO-based) and OnSIDES (MedDRA-based) were
evaluated using Best Match Average (BMA)~\cite{harispe2022semantic} with Resnik semantic
similarity~\cite{resnik1995using}. 

Resnik similarity uses ontological hierarchical structure and
information content:
\begin{equation}
\text{Resnik}(t_1, t_2) = \max_{t \in \text{CA}(t_1, t_2)} \text{IC}(t)
\end{equation}
where $\text{CA}(t_1, t_2)$ represents the common ancestors of terms
$t_1$ and $t_2$, and IC$(t)$ is the information content based on term
frequency in the dataset. BMA combines Resnik scores bidirectionally:
\begin{equation}
\small
\text{BMA}(D_1, D_2) = \frac{1}{2}\left(\frac{\sum_{t_1 \in D_1}
    \max_{t_2 \in D_2} \text{Res}(t_1, t_2)}{|D_1|} +
  \frac{\sum_{t_2 \in D_2} \max_{t_1 \in D_1} \text{Res}(t_1,
    t_2)}{|D_2|}\right)
\end{equation}
% \rh{ BMA and briefly explain what it does.}
%
For OnSIDES, we constructed the MedDRA hierarchy from {\em Preferred
  Terms}, {\em High Level Terms}, {\em High Level Group Terms}, and
{\em System Organ Class} levels. OnSIDES MedDRA codes were normalized
to Preferred Terms, and Resnik similarity was computed over the MedDRA
hierarchy using the same information content calculation approach as
SIDEKICK.

We computed similarity scores for all drug pairs and evaluated
classification performance using the Area Under the ROC (AUC-ROC),
with higher AUC indicating better discrimination between positive and
negative pairs. Table~\ref{tab:drug_target_results} present the comparative
results. SIDEKICK achieved an AUC of $0.7174$, representing 8.5\%
relative improvement over OnSIDES (AUC $0.6612$).
The superior performance of SIDEKICK is evident in the discrimination
between positive and negative pairs. SIDEKICK achieves a mean
similarity difference ($\Delta$) of 
% $\Delta$ of
$0.1711$, indicating 14\% better separation of true drug pairs. This
demonstrates that HPO's ontology structure captures mechanistically
relevant patterns more effectively than MedDRA's hierarchy.

\begin{table}[t]
\centering
\caption{Drug target prediction performance comparing SIDEKICK (HPO) and OnSIDES (MedDRA) using identical BMA+Resnik similarity. Mean($+$) and Mean($-$) represent average similarity scores for positive and negative pairs respectively. $\Delta$ shows the discrimination between classes.}
\label{tab:drug_target_results}
\begin{tabular}{lcccc}
\hline
\textbf{Method} & \textbf{AUC-ROC} & \textbf{Mean(+)} & \textbf{Mean(-)} & \textbf{$\Delta$} \\
\hline
OnSIDES (MedDRA+BMA+Resnik) & 0.6612 & 0.4412 & 0.2700 & 0.1711 \\
\textbf{SIDEKICK (HPO+BMA+Resnik)} & \textbf{0.7174} & \textbf{0.5935} & \textbf{0.3978} & \textbf{0.1957} \\
\hline
\end{tabular}
\end{table}

\subsubsection{Discussion}
The performance difference between SIDEKICK and OnSIDES, when both use
identical similarity methodology, demonstrates that ontology design
fundamentally impacts pharmacological applications. Although MedDRA
possesses a hierarchical structure, it was designed for regulatory
adverse event reporting with broad categorical groupings optimized for
safety surveillance rather than mechanistic reasoning. In contrast,
HPO was designed to ontologically structure the domain of abnormal
phenotypes in humans \cite{kohler2021human}, with axioms that explicitly relate to
potential mechanisms and anatomical structures affected in the
abnormality.

Our results validate both the quality of our HPO mappings and the
utility of HPO as a representational framework for adverse drug
reactions. SIDEKICK's improved performance for finding drugs with
shared mechanisms of action also shows it is useful for generating
hypotheses for new therapeutic indications, and can provide a
foundation for computational pharmacovigilance systems that reason
over drug safety profiles.

\subsection{Competency Question Evaluation}

To further validate SIDEKICK, we executed a series of Competency
Questions (CQs) that require traversing ontological hierarchies, and
answered them using SPARQL queries.  We evaluated two forms of
reasoning: (1) hierarchical traversal over the transitive closure of
\texttt{rdfs:subClassOf} relations in HPO and MONDO, and (2) federated
reasoning over ontology axioms that queries the logical definitions of
phenotypes and anatomical parthood relations stored in ontology
repositories such as
Ubergraph\footnote{\url{https://frink.renci.org/registry/kgs/ubergraph/}}.
We formulated questions that represent real-world pharmacovigilance
scenarios where a clinician searches for a general class of adverse
events (e.g., ``cardiovascular abnormalities'') rather than
enumerating hundreds of specific symptoms ({\em e.g.}, ``atrial
fibrillation'', ``myocardial infarction'', ``bradycardia'').

Table~\ref{tab:competency_questions} summarizes the results. The
evaluation demonstrates several reasoning capabilities. For anatomical
subsumption, the query for ``Abnormality of the cardiovascular
system'' (\verb_HP:0001626_) successfully retrieved 1,903 unique drugs. These
drugs were rarely annotated with this general term; rather, they were
linked to specific descendants such as ``Hypertension'' or
``Myocardial infarction'' via the inference $Drug \rightarrow
SpecificPhenotype \xrightarrow{is\_a} ParentPhenotype$. Regarding
physiological grouping, the query for ``Arrhythmia'' (\verb_HP:0011675_)
retrieves drugs based on physiological mechanisms rather than gross
anatomy, aggregating disjoint subtypes (e.g., Tachycardia,
Bradycardia) into a single axiomatic group without requiring manual
enumeration. We also demonstrated contextual safety reasoning by
constraining the relation type to \textit{contraindication},
successfully identifying drugs unsafe for patients with renal
conditions. This highlights the graph's ability to distinguish between
drugs that \textit{cause} a phenotype and those that are
\textit{contraindicated} by it using the same vocabulary. We further
utilized cross-ontology inference to query drugs indicated for
``Infectious Disease'' (\verb_MONDO:0005550_), leveraging the MONDO hierarchy
to aggregate drugs for specific infections.

Finally, we performed a federated query across SIDEKICK and
Ubergraph\footnote{\url{https://frink.renci.org/registry/kgs/ubergraph/}}
to demonstrate federated axiomatic reasoning, identifying drugs
affecting any anatomical part of the heart. This query utilizes the
logical axioms of HPO, which define phenotypes in terms of the
entities they affect (via \texttt{UPHENO:0000001}), and the transitive
parthood relations (\texttt{BFO:0000050}) in the UBERON anatomy
ontology. This allows SIDEKICK to retrieve drugs causing abnormalities
in specific heart structures (e.g., heart valves, ventricles) based on
implicit biological knowledge not present in the simple class
hierarchy. The federated query identified 504 drugs associated with 
abnormalities in 40 distinct cardiac anatomical structures, including 
myocardium, cardiac valves, and coronary vessels, yielding 2,676 
drug-structure associations. 

% For example, adalimumab was linked to 
% coronary artery atherosclerosis affecting the coronary artery, and 
% acetaminophen/hydrocodone to aortic regurgitation affecting the 
% aortic valve—associations derived through axiomatic reasoning rather 
% than explicit annotation.

\begin{table}[t]
\centering
\caption{Results of Competency Questions (CQs) demonstrating ontology-based reasoning. ``Unique Drugs'' indicates the number of distinct drug formulations identified.}
\label{tab:competency_questions}
\begin{tabular}{p{5.5cm}| p{3.5cm}| r}
\toprule
\textbf{Competency Question} & \textbf{Reasoning Type} & \textbf{Unique Drugs}  \\
\midrule
Find all drugs with any cardiovascular side effect & Anatomical Hierarchy & 1,903 \\
Find all drugs with any nervous system side effect & Anatomical Hierarchy & 1,907  \\
Find drugs affecting cardiac rhythm (Arrhythmia) & Physiological Grouping & 1,216 \\
Find drugs with metabolic side effects & Multi-level Hierarchy & 1,915  \\
Find drugs contraindicated in kidney conditions & Cross-System Safety & 184  \\
Find drugs indicated for any infectious disease & Disease Taxonomy & 434  \\
Find drugs affecting specific anatomical parts of heart & Federated Axiomatic & 504 \\
\bottomrule
\end{tabular}
\end{table}

\section{Conclusion}
We developed SIDEKICK, a knowledge graph of drug safety information
that addresses limitations in existing pharmacovigilance resources
through ontology-based standardization. By mapping drug side effects
to HPO, disease terms to MONDO, and drug interactions to RxNorm,
SIDEKICK provides the semantic richness essential for computational
reasoning in modern biomedical applications. % Our Graph-Retrieval
% Augmented Generation methodology achieves 96.7\% mapping accuracy,
% validated through expert evaluation of 980 side effect mappings,
% substantially outperforming traditional dictionary-based
% methods. 
Comparative evaluation against the state of the art OnSIDES database
demonstrates the practical impact of ontology-based representation;
SIDEKICK can improve the performance in predicting shared drug targets
based on side effects (ROC AUC increase of 8.5\%).
% SIDEKICK achieves an AUC of 0.7122 in predicting shared drug targets
% from side effect similarity, a 8.5\% relative improvement over
% OnSIDES. 
The knowledge graph, containing information extracted from over 50,000
drug labels, is serialized as RDF using the Semanticscience Integrated
Ontology as upper level ontology. It will interact with the Semantic
Web ecosystem and can be used to answer complex and federated SPARQL
queries.

\subsection{Limitations and future work}

Our mapping methodology achieved high accuracy; however, few ambiguous
terms and newly coined drug-specific terminology necessitated manual
disambiguation. Furthermore, we prioritized the extraction of binary
clinical associations; consequently, the current release lacks the
quantitative frequency data for adverse reactions available in
databases such as SIDER \cite{kuhn2016sider}. We also observed challenges in
normalizing terms related to patient populations within
contraindications to standard ontologies, which limits the ability to
reason over complex demographic exclusions. Regarding semantic
validation, the computational complexity of OWL 2 DL hindered full
validation of SIO axioms, necessitating a restriction to the EL subset
using the ELK reasoner. Finally, the resource currently relies
exclusively on FDA Structured Product Labels; therefore, it lacks
complementary signals for rare adverse events often derived from
post-marketing surveillance data, electronic health records, or
biomedical literature.

Future work prioritizes integrating genomic data via HPO's
disease--gene annotations to further improve pharmacogenomic
applications. Additionally, we aim to incorporate clinical trial
databases to validate drug repurposing hypotheses generated through
phenotype-based similarity.

\subsection{Availability}
SIDEKICK is freely available under the CC BY 4.0 International license. We deposited the
complete dataset at Zenodo (DOI: 10.5281/zenodo.17779317), and intend
to update the dataset yearly. Users can access the knowledge graph via
a web interface at \url{https://sidekick.bio2vec.net/} or query it
programmatically through the SPARQL endpoint at
\url{https://sidekick.bio2vec.net/sparql}, which includes example
queries. We provide source code, documentation, and tutorials at
\url{https://github.com/bio-ontology-research-group/sidekick/}.

\bibliographystyle{splncs04}
\bibliography{name}

\begin{thebibliography}{10}
\providecommand{\url}[1]{\texttt{#1}}
\providecommand{\urlprefix}{URL }
\providecommand{\doi}[1]{https://doi.org/#1}

\bibitem{alonso2021named}
Alonso~Casero, {\'A}.: Named entity recognition and normalization in biomedical literature: a practical case in SARS-CoV-2 literature. Ph.D. thesis, ETSI\_Informatica (2021)

\bibitem{antoniou2009web}
Antoniou, G., Harmelen, F.v.: Web ontology language: Owl. In: Handbook on ontologies, pp. 91--110. Springer (2009)

\bibitem{baker2005maintaining}
Baker, T.: Maintaining dublin core as a semantic web vocabulary. In: From Integrated Publication and Information Systems to Information and Knowledge Environments: Essays Dedicated to Erich J. Neuhold on the Occasion of His 65th Birthday, pp. 61--68. Springer (2005)

\bibitem{brown2004meddra}
Brown, E.G.: Using {MedDRA}: implications for risk management. Drug Safety  \textbf{27}(8),  591--602 (2004). \doi{10.2165/00002018-200427080-00010}

\bibitem{brown1999medical}
Brown, E.G., Wood, L., Wood, S.: The medical dictionary for regulatory activities (meddra). Drug safety  \textbf{20}(2),  109--117 (1999)

\bibitem{campillos2008drug}
Campillos, M., Kuhn, M., Gavin, A.C., Jensen, L.J., Bork, P.: Drug target identification using side-effect similarity. Science  \textbf{321}(5886),  263--266 (2008). \doi{10.1126/science.1158140}

\bibitem{classen1997adverse}
Classen, D.C., Pestotnik, S.L., Evans, R.S., Lloyd, J.F., Burke, J.P.: Adverse drug events in hospitalized patients: excess length of stay, extra costs, and attributable mortality. JAMA  \textbf{277}(4),  301--306 (1997). \doi{10.1001/jama.1997.03540280039031}

\bibitem{gene2019gene}
Consortium, G.O.: The gene ontology resource: 20 years and still going strong. Nucleic acids research  \textbf{47}(D1),  D330--D338 (2019)

\bibitem{dumontier2014sio}
Dumontier, M., Baker, C.J., Baran, J., Callahan, A., Chepelev, L., Cruz-Toledo, J., Del~Rio, N.R., Duck, G., Furlong, L.I., Keath, N., et~al.: The {Semanticscience Integrated Ontology (SIO)} for biomedical research and knowledge discovery. Journal of Biomedical Semantics  \textbf{5}(1),  1--11 (2014). \doi{10.1186/2041-1480-5-14}

\bibitem{gangemi2009ontology}
Gangemi, A., Presutti, V.: Ontology design patterns. In: Handbook on ontologies, pp. 221--243. Springer (2009)

\bibitem{gkoutos2018anatomy}
Gkoutos, G.V., Schofield, P.N., Hoehndorf, R.: The anatomy of phenotype ontologies: principles, properties and applications. Briefings in Bioinformatics  \textbf{19}(5),  1008--1021 (2018)

\bibitem{guarino1998formal}
Guarino, N.: Formal ontology in information systems: Proceedings of the first international conference (FOIS'98), June 6-8, Trento, Italy, vol.~46. IOS press (1998)

\bibitem{hagberg2008exploring}
Hagberg, A., Swart, P.J., Schult, D.A.: Exploring network structure, dynamics, and function using networkx. Tech. rep., Los Alamos National Laboratory (LANL), Los Alamos, NM (United States) (2008)

\bibitem{harispe2022semantic}
Harispe, S., Ranwez, S., Montmain, J., et~al.: Semantic similarity from natural language and ontology analysis. Springer Nature (2022)

\bibitem{hoehndorf2015role}
Hoehndorf, R., Schofield, P.N., Gkoutos, G.V.: The role of ontologies in biological and biomedical research: a functional perspective. Briefings in bioinformatics  \textbf{16}(6),  1069--1080 (2015)

\bibitem{kazakov2014incredible}
Kazakov, Y., Kr{\"o}tzsch, M., Siman{\v{c}}{\'\i}k, F.: The incredible elk: From polynomial procedures to efficient reasoning with $\mathcal{EL}$ ontologies. Journal of automated reasoning  \textbf{53}(1),  1--61 (2014)

\bibitem{kohler2021human}
K{\"o}hler, S., Gargano, M., Matentzoglu, N., Carmody, L.C., Lewis-Smith, D., Vasilevsky, N.A., Danis, D., Balagura, G., Baynam, G., Brower, A.M., et~al.: The {Human Phenotype Ontology} in 2021. Nucleic Acids Research  \textbf{49}(D1),  D1207--D1217 (2021). \doi{10.1093/nar/gkaa1043}

\bibitem{krech2002rdflib}
Krech, D., AAstrand~Grimnes, G., Higgins, G., Hees, J., Aucamp, I., Lindstr{\"o}m, N., Arndt, N., Sommer, A., Chuc, E., Herman, I., et~al.: Rdflib. Zenodo  (2002)

\bibitem{krotzsch2012owl}
Kr{\"o}tzsch, M.: Owl 2 profiles: An introduction to lightweight ontology languages. In: Reasoning Web International Summer School, pp. 112--183. Springer (2012)

\bibitem{kuhn2016sider}
Kuhn, M., Letunic, I., Jensen, L.J., Bork, P.: The {SIDER} database of drugs and side effects. Nucleic Acids Research  \textbf{44}(D1),  D1075--D1079 (2016). \doi{10.1093/nar/gkv1075}

\bibitem{kulmanov2021semantic}
Kulmanov, M., Smaili, F.Z., Gao, X., Hoehndorf, R.: Semantic similarity and machine learning with ontologies. Briefings in bioinformatics  \textbf{22}(4),  bbaa199 (2021)

\bibitem{mungall2010integrating}
Mungall, C.J., Gkoutos, G.V., Smith, C.L., Haendel, M.A., Lewis, S.E., Ashburner, M.: Integrating phenotype ontologies across multiple species. Genome biology  \textbf{11}(1), ~R2 (2010)

\bibitem{mungall2012uberon}
Mungall, C.J., Torniai, C., Gkoutos, G.V., Lewis, S.E., Haendel, M.A.: Uberon, an integrative multi-species anatomy ontology. Genome biology  \textbf{13}(1), ~R5 (2012)

\bibitem{noy2009bioportal}
Noy, N.F., Shah, N.H., Whetzel, P.L., Dai, B., Dorf, M., Griffith, N., Jonquet, C., Rubin, D.L., Storey, M.A., Chute, C.G., et~al.: Bioportal: ontologies and integrated data resources at the click of a mouse. Nucleic acids research  \textbf{37}(suppl\_2),  W170--W173 (2009)

\bibitem{nugent2016computational}
Nugent, T., Plachouras, V., Leidner, J.L.: Computational drug repositioning based on side-effects mined from social media. PeerJ Computer Science  \textbf{2}, ~e46 (2016). \doi{10.7717/peerj-cs.46}

\bibitem{pan2009resource}
Pan, J.Z.: Resource description framework. In: Handbook on ontologies, pp. 71--90. Springer (2009)

\bibitem{pesquita2009semantic}
Pesquita, C., Faria, D., Falcao, A.O., Lord, P., Couto, F.M.: Semantic similarity in biomedical ontologies. PLoS computational biology  \textbf{5}(7),  e1000443 (2009)

\bibitem{prud2014shape}
Prud'hommeaux, E., Labra~Gayo, J.E., Solbrig, H.: Shape expressions: an rdf validation and transformation language. In: Proceedings of the 10th International Conference on Semantic Systems. pp. 32--40 (2014)

\bibitem{pushpakom2019repurposing}
Pushpakom, S., Iorio, F., Eyers, P.A., Escott, K.J., Hopper, S., Wells, A., Doig, A., Guilliams, T., Latimer, J., McNamee, C., et~al.: Drug repurposing: progress, challenges and recommendations. Nature Reviews Drug Discovery  \textbf{18}(1),  41--58 (2019). \doi{10.1038/nrd.2018.168}

\bibitem{quinn1999hl7}
Quinn, J.: An hl7 (health level seven) overview. Journal of AHIMA  \textbf{70}(7),  32--4 (1999)

\bibitem{ratcliff1988pattern}
Ratcliff, J.W., Metzener, D.E., et~al.: Pattern matching: The gestalt approach. Dr. Dobb’s Journal  \textbf{13}(7), ~46 (1988)

\bibitem{resnik1995using}
Resnik, P.: Using information content to evaluate semantic similarity in a taxonomy. arXiv preprint cmp-lg/9511007  (1995)

\bibitem{smaili2020formal}
Smaili, F.Z., Gao, X., Hoehndorf, R.: Formal axioms in biomedical ontologies improve analysis and interpretation of associated data. Bioinformatics  \textbf{36}(7),  2229--2236 (2020)

\bibitem{tanaka2025onsides}
Tanaka, Y., Chen, H.Y., Belloni, P., Gisladottir, U., Kefeli, J., Patterson, J., Srinivasan, A., Zietz, M., Sirdeshmukh, G., Berkowitz, J., LaRow~Brown, K., Tatonetti, N.P.: {OnSIDES} database: extracting adverse drug events from drug labels using natural language processing models. Cell Reports Methods  \textbf{5}(4),  100698 (2025). \doi{10.1016/j.medj.2025.100642}

\bibitem{tatonetti2012offsides}
Tatonetti, N.P., Ye, P.P., Daneshjou, R., Altman, R.B.: Data-driven prediction of drug effects and interactions. Science Translational Medicine  \textbf{4}(125),  125ra31--125ra31 (2012). \doi{10.1126/scitranslmed.3003377}

\bibitem{vasilevsky2022mondo}
Vasilevsky, N.A., Matentzoglu, N., Toro, S., Flack, J.E., Hegde, H., Unni, D.R., Carmody, L.C., Duong, A., Munoz-Torres, M., Wilkinson, M.D., et~al.: Mondo: Unifying diseases for the world, by the world. medRxiv pp. 2022--04 (2022). \doi{10.1101/2022.04.13.22273750}

\bibitem{wilkinson2016fair}
Wilkinson, M.D., Dumontier, M., Aalbersberg, I.J., Appleton, G., Axton, M., Baak, A., Blomberg, N., Boiten, J.W., da~Silva~Santos, L.B., Bourne, P.E., et~al.: The fair guiding principles for scientific data management and stewardship. Scientific data  \textbf{3}(1), ~1--9 (2016)

\bibitem{wishart2018drugbank}
Wishart, D.S., Feunang, Y.D., Guo, A.C., Lo, E.J., Marcu, A., Grant, J.R., Sajed, T., Johnson, D., Li, C., Sayeeda, Z., et~al.: Drugbank 5.0: a major update to the drugbank database for 2018. Nucleic acids research  \textbf{46}(D1),  D1074--D1082 (2018)

\end{thebibliography}
\newpage
\appendix

\section{LLM Extraction Prompt}
\label{app:extraction_prompt}

The following prompt was used with Google Gemini 2.5 Flash to extract clinical entities from FDA Structured Product Labels:

\begin{quote}
\small
\texttt{You are an expert in extracting information from FDA drug labels. I have provided the text from a drug package label. Please extract the following information in a structured format:}

\texttt{IMPORTANT: Only respond with the extracted XML. Do not repeat any part of these instructions or the input text in your response.}

\texttt{1. Indications (what the drug is used for)}\\
\texttt{2. Contraindications (when the drug should not be used)}\\
\texttt{3. Side effects (with frequencies if available)}

\texttt{For each indication, contraindication, side effect, provide only one item per tag and include the exact line from the text that contains this information. Extract any and all indications, contraindications, side effects you find.}

\texttt{It is important to note that these side effects, indications and contraindications can be found in sections other than the ones specifically dedicated for them so search carefully across the entire text and find all of them.}

\texttt{Try to keep the indication, contraindication and side-effect names that you extract as short and straightforward as possible but accuracy is important.}

\texttt{Provide your response in the following XML format:}

\begin{verbatim}
<drug_information>
  <indications>
    <indication>
      <indication_name>INDICATION NAME</indication_name>
    </indication>
  </indications>
  <contraindications>
    <contraindication>
      <contraindication_name>CONTRAINDICATION NAME</contraindication_name>
    </contraindication>
  </contraindications>
  <side_effects>
    <side_effect>
      <side_effect_name>SIDE EFFECT NAME</side_effect_name>
    </side_effect>
  </side_effects>
</drug_information>
\end{verbatim}
\end{quote}

\section{Excluded LOINC Section Codes}
\label{app:loinc_blacklist}

During SPL text preprocessing, the following LOINC section codes were excluded as they do not contain clinically relevant information for extraction:

\begin{table}[h]
\centering
\small
\caption{LOINC section codes excluded from SPL text preprocessing}
\begin{tabular}{ll}
\toprule
\textbf{LOINC Code} & \textbf{Section Description} \\
\midrule
51945-4 & Package Label - Principal Display Panel \\
48780-1 & SPL Product Data Elements Section \\
44425-7 & Storage and Handling Section \\
34069-5 & How Supplied Section \\
34093-5 & References Section \\
59845-8 & Instructions for Use Section \\
42230-3 & SPL Patient Package Insert Section \\
42231-1 & SPL MedGuide Section \\
53413-1 & OTC - Questions Section \\
50565-1 & OTC - Keep Out of Reach of Children Section \\
34076-0 & Information for Patients Section \\
55106-9 & OTC - Active Ingredient Section \\
50569-3 & OTC - Ask Doctor Section \\
50568-5 & OTC - Ask Doctor/Pharmacist Section \\
34068-7 & Dosage \& Administration Section \\
51727-6 & Inactive Ingredient Section \\
55105-1 & OTC - Purpose Section \\
50570-1 & OTC - Do Not Use Section \\
50567-7 & OTC - When Using Section \\
50566-9 & OTC - Stop Use Section \\
53414-9 & OTC - Pregnancy or Breast Feeding Section \\
34090-1 & Clinical Pharmacology \\
43682-4 & Pharmacokinetics \\
43679-0 & Mechanism of Action \\
43681-6 & Pharmacodynamics \\
\bottomrule
\end{tabular}

\label{tab:loinc_blacklist}
\end{table}

\section{Entity Classification Categories}
\label{app:entity_classification}

Terms extracted from indication and contraindication sections were classified into seven categories using LLM-based classification:

\begin{enumerate}
    \item \textbf{Disease}: Medical conditions, disorders, syndromes (e.g., ``diabetes mellitus'', ``hypertension'', ``myocardial infarction'')
    
    \item \textbf{Phenotype}: Observable clinical signs, symptoms, or abnormalities (e.g., ``seizures'', ``hypotension'', ``bradycardia'')
    
    \item \textbf{Drug or Chemical}: Drug interactions, concomitant medications, or chemical substances (e.g., ``monoamine oxidase inhibitors'', ``warfarin'', ``alcohol'')
    
    \item \textbf{Allergy or Hypersensitivity}: Allergic reactions or hypersensitivity conditions. Terms automatically classified if containing keywords: ``hypersensitivity'', ``allergic'', or ``anaphylaxis''
    
    \item \textbf{Patient Population}: Demographics, life stages, or patient groups (e.g., ``pregnancy'', ``pediatric patients'', ``elderly'')
    
    \item \textbf{Procedure}: Medical or surgical procedures (e.g., ``surgery'', ``hemodialysis'', ``cardiac catheterization'')
    
    \item \textbf{Other}: Any terms that do not fit into the above categories
\end{enumerate}

Each category was mapped to appropriate standardized vocabularies: \textit{Disease} terms to MONDO Disease Ontology, \textit{Phenotype} terms to Human Phenotype Ontology (HPO), and \textit{Drug or Chemical} terms to RxNorm The unclassified terms were excluded in the final Knowledge Graph.

\section{API Configuration Parameters}
\label{app:api_parameters}

Table \ref{tab:api_params} summarizes the hyperparameters used for the Information extraction and Graph RAG ontology mapping process.

\begin{table}[h]
\centering
\caption{API configuration parameters for information extraction and ontology mapping}
\begin{tabular}{ll}
\toprule
\textbf{Parameter} & \textbf{Value} \\
\midrule
\multicolumn{2}{l}{\textit{Information Extraction}} \\
API Provider & OpenRouter \\
Model & google/gemini-2.5-flash \\
Temperature & 0.1 \\
Max Tokens & 50,000 \\
Batch Size & 500 documents \\
Sleep Between Batches & 10 seconds \\
Max Retries & 3 \\
Retry Delay & 5 seconds \\
\midrule
\multicolumn{2}{l}{\textit{Ontology Mapping (Graph-RAG)}} \\
API Provider & OpenRouter \\
Model & google/gemini-2.5-flash \\
Temperature & 0.1 \\
Max Tokens & 500 (mapping), 1,000 (side effects) \\
Batch Size & 1 term \\
Max Retries & 3 \\
Retry Delay & 2 seconds \\
Embedding Model & all-MiniLM-L6-v2 \\
Top-K Semantic Matches & 10 \\
Top-K Graph Nodes & 15 \\
\bottomrule
\end{tabular}

\label{tab:api_params}
\end{table}

\section{Example SPARQL Queries}
\label{app:sparql_queries}

This section provides representative SPARQL queries demonstrating SIDEKICK's ontology-based reasoning capabilities.

\subsection{Query 1: Cardiovascular Side Effects}

This query demonstrates hierarchical reasoning by finding all drugs with any cardiovascular-related side effects through transitive traversal of the HPO hierarchy:

\begin{lstlisting}[language=SPARQL,basicstyle=\small\ttfamily,breaklines=true]
PREFIX sio: <http://semanticscience.org/resource/>
PREFIX sk: <http://sidekick.bio2vec.net/>
PREFIX obo: <http://purl.obolibrary.org/obo/>
PREFIX rdfs: <http://www.w3.org/2000/01/rdf-schema#>

SELECT DISTINCT ?drug_name ?phenotype_label ?phenotype_id
WHERE {
    # Root term: Abnormality of the cardiovascular system
    BIND(obo:HP_0001626 AS ?cardiac_root)
    
    # Find all phenotypes that are subclasses (transitive)
    ?phenotype_id rdfs:subClassOf* ?cardiac_root .
    ?phenotype_id rdfs:label ?phenotype_label .
    
    # Find drugs with these side effects
    ?assoc a sk:SideEffect ;
           sk:refersToDrug ?collection ;
           sk:hasSideEffect ?phenotype_id .
    
    ?collection rdfs:label ?drug_name .
}
ORDER BY ?drug_name ?phenotype_label
\end{lstlisting}

\textbf{Result}: Identified 1,903 unique drugs with cardiovascular-related side effects, automatically aggregating specific conditions (e.g., hypertension, myocardial infarction, arrhythmia) without manual enumeration.

\subsection{Query 2: Renal Contraindications}

This query demonstrates cross-system reasoning by identifying drugs contraindicated in any kidney-related condition (combining phenotypes and diseases):

\begin{lstlisting}[language=SPARQL,basicstyle=\small\ttfamily,breaklines=true]
PREFIX sio: <http://semanticscience.org/resource/>
PREFIX sk: <http://sidekick.bio2vec.net/>
PREFIX obo: <http://purl.obolibrary.org/obo/>
PREFIX rdfs: <http://www.w3.org/2000/01/rdf-schema#>

SELECT DISTINCT ?drug_name ?contraindication_type ?phenotype_label
WHERE {
    # Root term: Abnormality of the kidney
    BIND(obo:HP_0000077 AS ?kidney_root)
    
    # Find kidney-related phenotypes
    ?phenotype_id rdfs:subClassOf* ?kidney_root .
    ?phenotype_id rdfs:label ?phenotype_label .
    
    # Phenotype contraindications
    {
        ?assoc a sk:PhenotypeContraindication ;
               sk:refersToDrug ?collection ;
               sk:isContraindicatedInPhenotype ?phenotype_id .
        BIND("Phenotype" AS ?contraindication_type)
    }
    UNION
    {
        # Disease contraindications with "renal" or "kidney"
        ?assoc a sk:DiseaseContraindication ;
               sk:refersToDrug ?collection ;
               sk:isContraindicatedInDisease ?disease_id .
        BIND("Disease" AS ?contraindication_type)
        ?disease_id rdfs:label ?phenotype_label .
        FILTER(CONTAINS(LCASE(?phenotype_label), "renal") || 
               CONTAINS(LCASE(?phenotype_label), "kidney"))
    }
    
    ?collection rdfs:label ?drug_name .
}
ORDER BY ?drug_name
\end{lstlisting}

\textbf{Result}: Identified 184 drugs contraindicated in kidney-related conditions, demonstrating the ability to distinguish drugs that \textit{cause} kidney problems from those that are \textit{contraindicated} in existing kidney disease.

\subsection{Query 3: Infectious Disease Indications}

This query demonstrates disease taxonomy reasoning using the MONDO Disease Ontology:

\begin{lstlisting}[language=SPARQL,basicstyle=\small\ttfamily,breaklines=true]
PREFIX sio: <http://semanticscience.org/resource/>
PREFIX sk: <http://sidekick.bio2vec.net/>
PREFIX obo: <http://purl.obolibrary.org/obo/>
PREFIX rdfs: <http://www.w3.org/2000/01/rdf-schema#>

SELECT DISTINCT ?drug_name ?disease_label ?disease_id
WHERE {
    # Root term: Infectious disease (MONDO)
    BIND(obo:MONDO_0005550 AS ?infectious_root)
    
    # Find all infectious diseases via hierarchy
    ?disease_id rdfs:subClassOf* ?infectious_root .
    ?disease_id rdfs:label ?disease_label .
    
    # Find drugs indicated for these diseases
    ?assoc a sk:DiseaseIndication ;
           sk:refersToDrug ?collection ;
           sk:isIndicatedForDisease ?disease_id .
    
    ?collection rdfs:label ?drug_name .
}
ORDER BY ?drug_name
\end{lstlisting}

\textbf{Result}: Identified 434 drugs indicated for infectious diseases, automatically aggregating antimicrobial, antiviral, and antifungal agents through disease taxonomy reasoning.

\subsection{Query 4: Federated Axiomatic Reasoning}

This query demonstrates federated reasoning across SIDEKICK and the Ubergraph knowledge graph, utilizing HPO's logical axioms that link phenotypes to anatomical structures via UBERON:

\begin{lstlisting}[language=SPARQL,basicstyle=\small\ttfamily,breaklines=true]
PREFIX sio: <http://semanticscience.org/resource/>
PREFIX sk: <http://sidekick.bio2vec.net/>
PREFIX obo: <http://purl.obolibrary.org/obo/>
PREFIX rdfs: <http://www.w3.org/2000/01/rdf-schema#>

SELECT DISTINCT ?drug_name ?phenotype_label ?heart_part_label
WHERE {
    # Local query: Find drugs and side effects
    ?assoc a sk:SideEffect ;
           sk:refersToDrug ?collection ;
           sk:hasSideEffect ?phenotype_id .
    
    ?collection rdfs:label ?drug_name .
    ?phenotype_id rdfs:label ?phenotype_label .
    
    # Federated query: Find anatomical parts
    SERVICE <https://ubergraph.apps.renci.org/sparql> {
        # What entity does this phenotype affect?
        ?phenotype_id obo:UPHENO_0000001 ?affected_entity .
        
        # Is it part of the heart (transitively)?
        ?affected_entity obo:BFO_0000050* obo:UBERON_0000948 .
        
        # Get the anatomical part label
        ?affected_entity rdfs:label ?heart_part_label .
        
        FILTER(?affected_entity != obo:UBERON_0000948)
    }
}
ORDER BY ?drug_name ?heart_part_label
\end{lstlisting}

\textbf{Result}: Identified 504 drugs associated with abnormalities in 40 distinct cardiac anatomical structures (e.g., myocardium, cardiac valves, coronary vessels), yielding 2,676 drug-structure associations through axiomatic reasoning not present in the simple class hierarchy.

\end{document}